\documentclass[twocolumn,showpacs,superscriptaddress]{revtex4-1}
\usepackage{graphicx}
\usepackage{amsmath}
\usepackage{amssymb}
\usepackage{color}
\usepackage{wasysym}
\usepackage{enumerate}
\usepackage[colorlinks]{hyperref}
\usepackage{ifthen}
\usepackage{xspace}

\newcommand{\tw}{\ensuremath{t_\mathrm{w}}\xspace}
\newcommand{\Tc}{\ensuremath{T_\mathrm{c}}\xspace}
\newcommand{\NRep}{\ensuremath{N_{\text{Rep}}}\xspace}
\newcommand{\NS}{\ensuremath{N_{\text{S}}}\xspace}

\begin{document}

\title{Temperature chaos is present in off-equilibrium spin-glass dynamics}

\author{Marco~Baity-Jesi}\affiliation{Eawag,  \"Uberlandstrasse 133, CH-8600 D\"ubendorf, Switzerland}

\author{Enrico~Calore}\affiliation{Dipartimento di Fisica e Scienze della
  Terra, Universit\`a di Ferrara e INFN, Sezione di Ferrara, I-44122
  Ferrara, Italy}

\author{Andr\'es~Cruz}\affiliation{Departamento de F\'\i{}sica Te\'orica,
  Universidad de Zaragoza, 50009 Zaragoza,
  Spain}\affiliation{Instituto de Biocomputaci\'on y F\'{\i}sica de
  Sistemas Complejos (BIFI), 50018 Zaragoza, Spain}

\author{Luis Antonio~Fernandez}\affiliation{Departamento de F\'\i{}sica
  Te\'orica, Universidad Complutense, 28040 Madrid,
  Spain}\affiliation{Instituto de Biocomputaci\'on y F\'{\i}sica de
  Sistemas Complejos (BIFI), 50018 Zaragoza, Spain}

\author{Jos\'e Miguel~Gil-Narvion}\affiliation{Instituto de Biocomputaci\'on y
  F\'{\i}sica de Sistemas Complejos (BIFI), 50018 Zaragoza, Spain}

\author{Isidoro~Gonzalez-Adalid Pemartin}\affiliation{Departamento  de F\'\i{}sica Te\'orica, Universidad Complutense, 28040 Madrid, Spain}

\author{Antonio~Gordillo-Guerrero}\affiliation{Departamento de
  Ingenier\'{\i}a El\'ectrica, Electr\'onica y Autom\'atica, U. de
  Extremadura, 10003, C\'aceres, Spain}\affiliation{Instituto de
  Computaci\'on Cient\'{\i}fica Avanzada (ICCAEx), Universidad de
  Extremadura, 06006 Badajoz, Spain}\affiliation{Instituto de
  Biocomputaci\'on y F\'{\i}sica de Sistemas Complejos (BIFI), 50018
  Zaragoza, Spain}

\author{David~I\~niguez}\affiliation{Instituto de Biocomputaci\'on y
  F\'{\i}sica de Sistemas Complejos (BIFI), 50018 Zaragoza,
  Spain}\affiliation{Fundaci\'on ARAID, Diputaci\'on General de
  Arag\'on, 50018 Zaragoza, Spain}\affiliation{Departamento de F\'\i{}sica Te\'orica,
  Universidad de Zaragoza, 50009 Zaragoza, Spain}

\author{Andrea~Maiorano}\affiliation{Dipartimento di Biotecnologie, Chimica e
  Farmacia, Universit\`a degli studi di Siena, 53100, Siena,
  Italy}\affiliation{INFN, Sezione di Roma 1, I-00185 Rome,
  Italy}\affiliation{Instituto de Biocomputaci\'on y F\'{\i}sica de Sistemas
Complejos (BIFI), 50018 Zaragoza, Spain}

\author{Enzo~Marinari}\affiliation{Dipartimento di Fisica, Sapienza
  Universit\`a di Roma, and CNR-Nanotec,
  I-00185 Rome, Italy}\affiliation{INFN, Sezione di Roma 1, I-00185 Rome,
  Italy}

\author{V\'ictor~Martin-Mayor}\affiliation{Departamento de F\'\i{}sica
  Te\'orica, Universidad Complutense, 28040 Madrid,
  Spain}\affiliation{Instituto de Biocomputaci\'on y F\'{\i}sica de
  Sistemas Complejos (BIFI), 50018 Zaragoza, Spain}

\author{Javier~Moreno-Gordo}\email{ORCID:0000-0002-0420-8605 \,\,  jmorenogordo@gmail.com}\affiliation{Instituto de Biocomputaci\'on y
  F\'{\i}sica de Sistemas Complejos (BIFI), 50018 Zaragoza,
  Spain}\affiliation{Departamento de F\'\i{}sica Te\'orica,
  Universidad de Zaragoza, 50009 Zaragoza, Spain}

\author{Antonio~Mu\~noz-Sudupe}\affiliation{Departamento de F\'\i{}sica
  Te\'orica, Universidad Complutense, 28040 Madrid,
  Spain}\affiliation{Instituto de Biocomputaci\'on y F\'{\i}sica de
  Sistemas Complejos (BIFI), 50018 Zaragoza, Spain}

\author{Denis~Navarro}\affiliation{Departamento de Ingenier\'{\i}a,
  Electr\'onica y Comunicaciones and I3A, U. de Zaragoza, 50018
  Zaragoza, Spain}

\author{Ilaria~Paga} \affiliation{Dipartimento di Fisica, Sapienza Universit\`a di Roma,
  INFN, Sezione di Roma 1, I-00185 Rome,Italy} \affiliation{Departamento de
  F\'\i{}sica Te\'orica, Universidad Complutense, 28040 Madrid, Spain}

\author{Giorgio~Parisi}\affiliation{Dipartimento di Fisica, Sapienza
  Universit\`a di Roma, and CNR-Nanotec,
  I-00185 Rome, Italy}\affiliation{INFN, Sezione di Roma 1, I-00185 Rome,
  Italy}

\author{Sergio~Perez-Gaviro}\affiliation{Centro Universitario de la Defensa, 50090 Zaragoza, Spain}\affiliation{Instituto de Biocomputaci\'on y F\'{\i}sica de Sistemas
  Complejos (BIFI), 50018 Zaragoza, Spain}\affiliation{Departamento de
  F\'\i{}sica Te\'orica, Universidad de Zaragoza, 50009 Zaragoza, Spain}

\author{Federico~Ricci-Tersenghi}\affiliation{Dipartimento di Fisica, Sapienza
  Universit\`a di Roma, and CNR-Nanotec,
  I-00185 Rome, Italy}\affiliation{INFN, Sezione di Roma 1, I-00185 Rome,
  Italy}

\author{Juan Jes\'us~Ruiz-Lorenzo}\affiliation{Departamento de F\'{\i}sica,
  Universidad de Extremadura, 06006 Badajoz,
  Spain}\affiliation{Instituto de Computaci\'on Cient\'{\i}fica
  Avanzada (ICCAEx), Universidad de Extremadura, 06006 Badajoz,
  Spain}\affiliation{Instituto de Biocomputaci\'on y F\'{\i}sica de
  Sistemas Complejos (BIFI), 50018 Zaragoza, Spain}

\author{Sebastiano Fabio~Schifano}\affiliation{Dipartimento di Scienze Chimiche e Farmaceutiche, Universit\`a di Ferrara e INFN  Sezione di Ferrara, I-44122 Ferrara, Italy}

\author{Beatriz~Seoane}\affiliation{Departamento de F\'\i{}sica
  Te\'orica, Universidad Complutense, 28040 Madrid,
  Spain}\affiliation{Instituto de Biocomputaci\'on y F\'{\i}sica de
  Sistemas Complejos (BIFI), 50018 Zaragoza, Spain}

\author{Alfonso~Tarancon}\affiliation{Departamento de F\'\i{}sica
  Te\'orica, Universidad de Zaragoza, 50009 Zaragoza,
  Spain}\affiliation{Instituto de Biocomputaci\'on y F\'{\i}sica de
  Sistemas Complejos (BIFI), 50018 Zaragoza, Spain}

\author{Raffaele~Tripiccione}\affiliation{Dipartimento di Fisica e Scienze
  della Terra, Universit\`a di Ferrara e INFN, Sezione di Ferrara,
  I-44122 Ferrara, Italy}

\author{David~Yllanes}\affiliation{Chan Zuckerberg Biohub, San Francisco, CA, 94158, United States}
\affiliation{Instituto de Biocomputaci\'on y F\'{\i}sica de
  Sistemas Complejos (BIFI), 50018 Zaragoza, Spain}

\collaboration{Janus Collaboration}

\date{\today}

\begin{abstract}
{\large\bfseries\noindent Abstract}\\ Experiments featuring non-equilibrium
glassy dynamics under temperature changes still await interpretation. There is
a widespread feeling that temperature chaos (an extreme sensitivity of the
glass to temperature changes) should play a major role but, up to now,
this phenomenon has been investigated solely under
equilibrium conditions. In fact, the very existence of a chaotic effect in the
non-equilibrium dynamics is yet to be established. In this article, we tackle
this problem through a large simulation of the 3D Edwards-Anderson model,
carried out on the Janus II supercomputer. We find a
dynamic effect that closely parallels equilibrium temperature chaos. This
dynamic temperature-chaos effect is spatially heterogeneous to a large degree
and turns out to be controlled by the spin-glass coherence length $\xi$.
Indeed, an emerging length-scale $\xi^*$ rules the crossover from weak (at $\xi
\ll \xi^*$) to strong chaos ($\xi \gg \xi^*$). Extrapolations of $\xi^*$ to
relevant experimental conditions are provided.  \end{abstract}

\maketitle


\section*{Introduction} 
An important lesson taught by spin
glasses~\cite{young:98} regards the fragility of the glassy phase in response
to perturbations such as changes in temperature ---temperature chaos
(TC)~\cite{mckay:82,bray:87b,banavar:87,kondor:89,ney-nifle:97,ney-nifle:98,billoire:00,mulet:01,billoire:02,krzakala:02,rizzo:03,sasaki:05,katzgraber:07,parisi:10,fernandez:13,billoire:14,wang:15,billoire:18}---,
in the couplings~\cite{ney-nifle:97,ney-nifle:98,sasaki:05,katzgraber:07} or
in the external magnetic field~\cite{kondor:89,ritort:94,billoire:03}. In particular,
it is somewhat
controversial~\cite{komori:00,berthier:02,picco:01,takayama:02,maiorano:05,jimenez:05}
whether or not TC is the physical mechanism underlying the spectacular
rejuvenation and memory effects found in spin
glasses~\cite{jonason:98,lundgren:83,jonsson:00,hammann:00} and several other
materials~\cite{ozon:03,bellon:00,yardimci:03,bouchaud:01b,mueller:04}.
Indeed, a major obstacle in the analysis of these non-equilibrium experiments
is that TC is a theoretical notion which is solely defined in an equilibrium
context.

Specifically, TC means that the spin configurations that are typical
from the Boltzmann weight at temperature $T_1$ are very atypical at
temperature $T_2$ (no matter how close the two temperatures $T_1$ and
$T_2$ are).

This equilibrum notion of TC has turned out to be
  remarkably elusive, even in the context of Mean-Field models
  (i.e., models that can be solved exactly in the  Mean-Field
  approximation). Indeed, establishing the existence of TC for the
  Sherrington-Kirkpatrick model has been a real tour de
    force~\cite{rizzo:03}. Although Sherrington-Kirkpatrick's model
  is the Mean-Field model of more direct relevance for this work, let
  us recall for completeness that TC has been investigated as
  well in other Mean-Field systems named $p$-spin models. In these
  models, groups of $p\geq 3$ spins interact (instead, $p=2$ for
  Sherrington-Kirkpatrick). Surprisingly enough, one finds different
  behaviors.  On the one hand, we have a recent mathematical proof of
  the absence of TC in the homogeneous spherical $p$-spin model~\cite{subag:17}, in
  agreement with a previous claim based on physical
  arguments~\cite{kurchan:93}. On the other hand, TC should be
  expected when one mixes several values of $p$~\cite{barrat:97}, as
  confirmed by a quite recent mathematical
  analysis~\cite{chen:14,panchenko:16,chen:17,arous:20}.
  Unfortunately, the mathematically rigorous analysis of TC in
  off-equilibrium dynamics seems out of reach for now, even in the Mean-Field
  context.

In order to obtain experimentally relevant results, one needs to go beyond the Mean-Field approximation and study short-range spin glasses, represented by the Edwards-Anderson model~\cite{edwards:75,edwards:76}. In this case, analytical investigations are even more difficult, but the equilibrium notion of TC that we have outlined above has been studied through numerical simulations. Yet, these equilibrium simulations have been limited to small system sizes by the severe dynamic slowing down~\cite{ney-nifle:97,ney-nifle:98,billoire:00,
krzakala:02,sasaki:05,katzgraber:07,fernandez:13,billoire:14,wang:15,billoire:18}.
 
Here we tackle the problem from a different approach by showing that a non-equilibrium TC effect is indeed present in the dynamics of a large spin-glass sample in three spatial dimensions (our simulations of the Edwards-Anderson model are carried out on the Janus II custom-built supercomputer~\cite{janus:14}). In a reincarnation of the statics-dynamics equivalence principle~\cite{barrat:01,janus:08b,janus:10b,janus:16}, just as
equilibrium TC is ruled by the system size, dynamic TC is found to be governed
by the time-growing spin-glass coherence length $\xi(\tw)$, where the waiting
time $\tw$ is the time elapsed since the system was suddenly quenched from
some very high temperature to the working temperature $T$. Below the
  critical temperature, $T<\Tc$, the spin glass is perennially
  out of equilibrium as evinced by the never-ending (and sluggish) growth of
  glassy magnetic domains of size $\xi(\tw)$, see
  Refs.~\cite{janus:18,zhai:19} for instance. Now, the extreme sample-to-sample variations
found in small equilibrated
systems~\cite{fernandez:13,billoire:11,billoire:14,martin-mayor:15,fernandez:16,
  billoire:18} translate into a strong spatial heterogeneity of dynamic TC.
Despite such strong fluctuations, our large-scale simulations allow us to
observe traces of the effect even in averages over the whole system.  In close
analogy with equilibrium studies~\cite{fernandez:13}, however, dynamic TC can
only be fully understood through a statistical analysis of the spatial
heterogeneity.  A crossover length $\xi^*$ emerges such that TC becomes
sizeable only when $\xi(\tw)>\xi^*$. We find that $\xi^*$ diverges when the
two observation temperatures $T_1$ and $T_2$ approach. The analysis of this
divergence reveals that $\xi^*$ is the non-equilibrium partner of the
equilibrium chaotic length~\cite{fisher:86,bray:87b}. The large values of
$\xi(\tw)$ that we reach with Janus~II allow us to perform mild extrapolations
to reach the most recent experimental regime~\cite{zhai:20b}.

In equilibrium, sample-averaged signals of TC become more visible when the size
of the system increases~\cite{fernandez:13}.  Analogously, off-equilibrium 
a weak chaotic effect grows with $\xi(\tw)$  when the whole system is considered on
average.  Hoping that studying spatial heterogeneities will help us to unveil
dynamic TC, we shall consider spatial regions of spherical shape and
linear size $\sim \xi(\tw)$, chosen randomly within a very large spin glass.
Statics-dynamics equivalence suggests regarding these spheres as the
non-equilibrium analogue of small equilibrated samples of linear size $\sim
\xi(\tw)$. The analogy with equilibrium
studies~\cite{fernandez:13,billoire:14,billoire:18} suggests that a small
fraction of our spheres will display strong TC. The important question will be
how this rare-event phenomenon evolves as $\xi(\tw)$ grows (in equilibrium, the
fraction of samples not displaying TC is expected to diminish
exponentially with the number of spins contained in the
sample~\cite{rizzo:03,parisi:10}).
\section*{Results and Discussion}
{\noindent \bfseries Model.} We simulate the standard Edwards-Anderson model in a
three-dimensional cubic lattice of linear size $L=160$ and periodic boundary
conditions. In each lattice node $\boldsymbol{x}$, we place an Ising spin
($S_{\boldsymbol{x}}=\pm 1$). Lattice nearest-neighbors spins interact through the
Hamiltonian
$H = - \sum_{\langle \boldsymbol{x},\boldsymbol{y}\rangle}
J_{\boldsymbol{x}\boldsymbol{y}}S_{\boldsymbol{x}} S_{\boldsymbol{y}}\,.$ The couplings
$J_{\boldsymbol{x}\boldsymbol{y}}$ are independent and identically distributed random
variables ($J_{\boldsymbol{x}\boldsymbol{y}}=\pm 1$ with $1/2$ probability), fixed
when the simulation starts (quenched disorder). This model
exhibits a spin-glass transition at temperature
$T_c = 1.1019(29)$~\cite{janus:13}. We refer to each
realization of the couplings as a sample. Statistically independent simulations of a given sample are named replicas. We have considerably extended the simulation of Baity et. al.\cite{janus:18}, by simulating $\NRep=512$ replicas (rather than 256) of the same $\NS = 16$ samples considered in~\cite{janus:18}, in the
temperature range $0.625\leq T \leq 1.1$. 

We simulate the non-equilibrium dynamics with a Metropolis algorithm. In this
way, one picosecond of physical time roughly corresponds to a full-lattice
Metropolis sweep. At the initial time $\tw=0$ the spin configuration is fully
random (i.e., we quench from infinite temperature).  The subsequent growth of
spin-glass domains is characterized by the spin-glass coherence length
$\xi(\tw)$. Specifically, we use the $\xi_{1,2}$ integral estimators, see
Refs.~\cite{fernandez:19,janus:18,janus:08b,janus:09b} for details [the
  main steps in the computation of $\xi_{1,2}$ are also sketched in
  Eqs.~(\ref{eq:def_corr_func_2T},\ref{eq:def_integral_2T},\ref{eq:def_correlation_length_2T}),
  where one should set $T_1=T_2=T$].

Finally, let us briefly comment on our choices for $\NRep$ and $\NS$.  A
  detailed analysis~\cite{janus:18,fernandez:19b} shows that, for a given
  total numerical effort $\NS\times \NRep$, errors in $\xi$ are minimized if
  $\NRep\gg\NS$. Furthermore, Supplementary Note 1 shows that having
  $\NRep\gg 1$ is crucial as well for the main quantities considered in this
  work (see definitions below). Therefore, given our finite computational
  resources, we have chosen to limit ourselves to $\NS = 16$. This small
  number of samples is partly compensated by the fact that we are working
  close to the experimental regime $L\gg\xi$ [we remark that $\NS=1$ in
  typical experiments: indeed, statics-dynamics equivalence suggests that the
  number of statistically independent events is proportional to
  $\NS(L/\xi)^3$].
\\

{\noindent \bfseries The local chaotic parameter.} We shall compare
the spin textures from temperature $T_1$ and waiting time $t_{\mathrm{w}1}$
with those from temperature $T_2$ and waiting time $t_{\mathrm{w}2}$ (we
consider $T_1\leq T_2\leq\Tc$). A fair comparison requires 
that the two configurations be ordered at the
same lengthscale, which we ensure by imposing the condition
\begin{equation}\label{eq:el_reloj_doble}
\xi(t_{\mathrm{w}1},T_1)=\xi(t_{\mathrm{w}2},T_2)=\xi\,.
\end{equation}

\begin{figure}[t] \centering \includegraphics[width=0.48\textwidth]{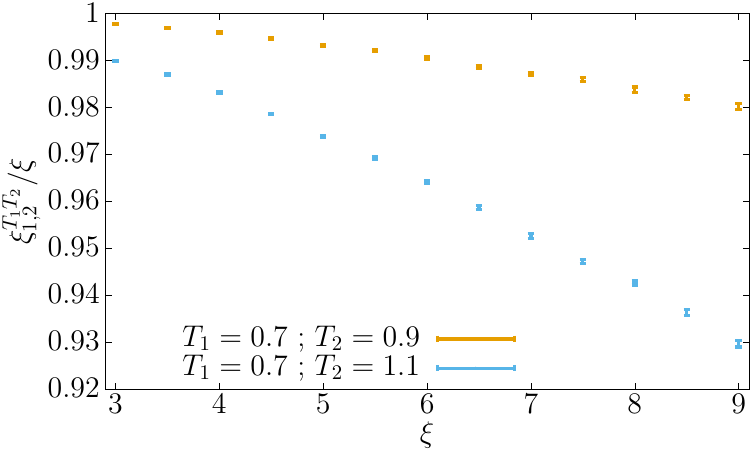}
  \caption{{\bf Non-equilibrium temperature chaos is weak when averaging over
the whole system.} We compare typical spin configurations at temperature $T_1$
and time $t_{\mathrm{w}1}$ with configurations at $T_2$ and time
$t_{\mathrm{w}2}$. The comparison is carried through a global estimator of the
coherence length of their overlap $\xi^{T_1T_2}_{1,2}$ which represents the
maximum lengthscale at which configurations at temperatures $T_1$ and $T_2$
still look similar, see Methods section for further details. The two times
$t_{\mathrm{w}1}$ and $t_{\mathrm{w}2}$ are chosen in such a way that the
configurations at both temperatures have glassy-domains of the same size,
namely $\xi_{1,2}(t_{\mathrm{w}1},T_1)=\xi_{1,2}(t_{\mathrm{w}2},T_2)=\xi$.
The figure shows the ratio $\xi^{T_1T_2}_{1,2}/\xi$ as a function of $\xi$ for
two pairs of temperatures $(T_1,T_2)$, recall that $\Tc\approx
1.1$~\cite{janus:13}---. Under the hypothesis of fully developed Temperature Chaos, we would
expect $\xi^{T_1T_2}_{1,2}$ to be negligible compared to $\xi$. Instead, our
data shows only a small decrease of $\xi^{T_1T_2}_{1,2}/\xi$ with growing $\xi$
(the larger the difference $T_2-T_1$ the more pronounced the decrease). Error bars represent one standard deviation.}
 \label{fig:xi_T1T2}
\end{figure}

A first investigation of TC is shown in Fig.~\ref{fig:xi_T1T2}.
The overlap, computed over the whole sample, of two systems 
satisfying condition Eq.~\eqref{eq:el_reloj_doble} is used 
to search for a coarse-grained chaotic effect. The resulting
signal is measurable but weak. Instead, as explained in the 
introduction, spin configurations should be compared locally. Specifically,
we consider spherical regions.  We start by choosing $N_{\mathrm{sph}}=8000$
centers for the spheres on each sample. The spheres' centers are chosen
randomly, with uniform probability, on the dual lattice which, in a cubic lattice with periodic boundary conditions, is another cubic lattice of the same size, also periodic boundary condition. The nodes of the dual lattice are the centers of the elementary cells of the original lattice. The radii of the spheres are varied, but their
centers are held fixed. Let $B_{s,r}$ be the $s$-th ball of radius $r$. Our
basic observable is the overlap between replica $\sigma$ (at temperature
$T_1$), and replica $\tau\neq\sigma$ (at temperature $T_2$):
\begin{equation}\label{eq:sphere_overlap}
q_{T_1,T_2}^{s,r;\sigma,\tau}(\xi) = \dfrac{1}{N_r} \sum_{\boldsymbol{x}\in B_{s,r}} s_{\boldsymbol{x}}^{\sigma,T_1}(t_{\mathrm{w}1}) s_{\boldsymbol{x}}^{\tau,T_2}(t_{\mathrm{w}2}) \>\> ,
\end{equation}
where $N_r$ is the number of spins in the ball, and $t_{\mathrm{w}1}$ and
$t_{\mathrm{w}2}$ are chosen according to Eq.~\eqref{eq:el_reloj_doble}.
Averages over thermal histories, indicated by $\langle\ldots\rangle_T$, are
computed by averaging over $\sigma$ and $\tau$.

\begin{figure}[b] \centering \includegraphics[width=0.48\textwidth]{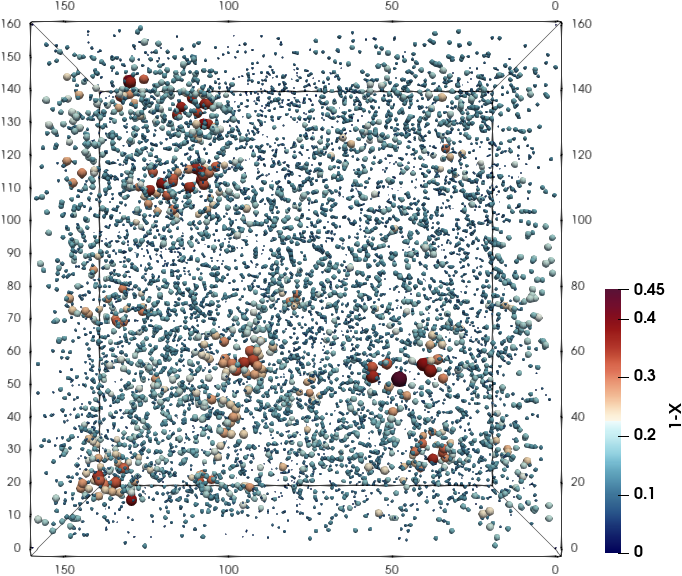}
  \caption{{\bf Dynamic temperature chaos is spatially heterogeneous.} The
    8000 randomly chosen spheres in a sample of size $L=160$ are depicted with a
    color code depending on $1-X$ [$X$ is the chaotic parameter,
    Eq.~\eqref{eq:def_chaotic_parameter}, as computed for spheres of radius
    $r=12$, $\xi=12$ and temperatures $T_1=0.7$ and $T_2=1.0$]. For
    visualization purposes, spheres are represented with a radius $12(1-X)$, so
    that only fully chaotic spheres (i.e., X=0) have their real size.}
  \label{fig:heterogeneidad}
\end{figure}

Next, we generalize the so-called
chaotic parameter~\cite{ritort:94,ney-nifle:97,fernandez:13,billoire:14} as
\begin{equation} 
  X^{s,r}_{T_1,T_2}(\xi) = \dfrac{\langle [q_{T_1,T_2}^{s,r;\sigma,\tau}(\xi)]^2\rangle_T}{\sqrt{\langle[q_{T_1,T_1}^{s,r;\sigma,\tau}(\xi)]^2\rangle_T{ \,\langle[q_{T_2,T_2}^{s,r;\sigma,\tau}(\xi)]^2\rangle_T}}}\>\>, \label{eq:def_chaotic_parameter}
\end{equation}
The extremal values of the chaotic parameter have a simple interpretation:
$X^{s,r}_{T_1,T_2}=1$ corresponds with a situation in which spin
configurations in the ball $B_{s,r}$, at temperatures $T_1$ and $T_2$, are
completely indistinguishable (absence of chaos) while $X^{s,r}_{T_1,T_2}=0$
corresponds to completely different configurations (strong TC). A representative example our results is shown in Fig.~\ref{fig:heterogeneidad}.

Our main focus will be on the distribution function
$F(X,T_1,T_2,\xi,r)=\text{Probability}[X^{s,r}_{T_1,T_2}(\xi)<X]$ and on its
inverse $X(F,T_1,T_2,\xi,r)$.

\begin{figure}[t]
  \centering
   \includegraphics[width=0.49\textwidth]{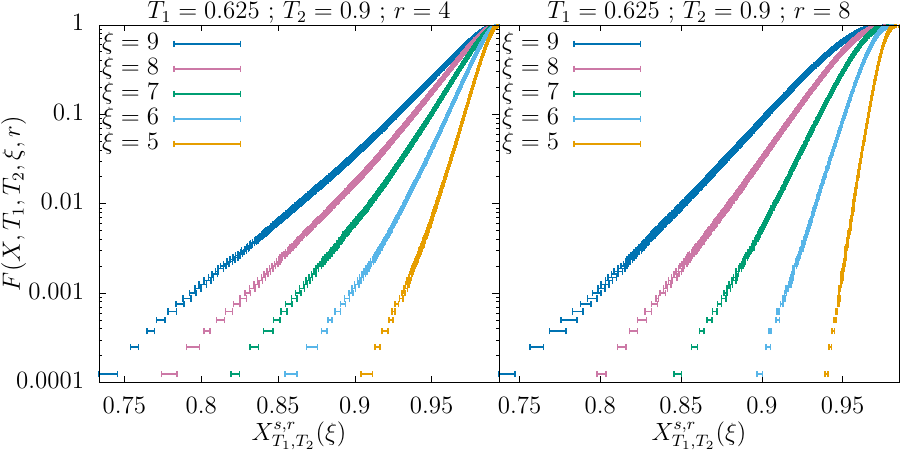}
   \caption{{\bf Temperature chaos increases with the coherence length.} The figure
     shows the distribution function $F(X,T_1,T_2,\xi,r)$ for temperatures $T_1=0.625$ and
     $T_2=0.9$, for spheres of radius $r=4$ and $r=8$, as
     computed for various values of $\xi$. The distributions have been
     extrapolated to infinite number of replicas $\NRep=\infty$, see Supplementary Note 1 for further details. Error bars, that represent one standard deviation, are
     horizontal, because we have actually extrapolated the chaotic parameter, which is its inverse function
     $X(F,T_1,T_2,\xi,r)$. Most of the spheres have a chaotic parameter very
     close to $X=1$ (absence of chaos). However, if we fix our attention, for instance, on
     percentile 1 (i.e., $F=0.01$) we see that the corresponding value of $X$
     decreases monotonically (and significantly) as $\xi$ grows,
     signaling a developing chaotic effect. This trend
     is clear both for spheres of radius $r=4$ and $r=8$.}
 \label{fig:distribution_function_xi}
\end{figure}

{\noindent \bfseries The rare-event analysis.} Representative examples of distribution
functions $F(X,T_1,T_2,\xi,r)$ are shown in
Fig.~\ref{fig:distribution_function_xi}. We see that, in close analogy with
equilibrium systems~\cite{fernandez:13,billoire:14,billoire:18}, while most
spheres exhibit a very weak TC ($X>0.9$, say), there is a fraction of
spheres displaying smaller $X$ (stronger chaos). Note that
the probability $F$ of finding spheres with $X$ smaller
than any prefixed value increases when $\xi$ grows.

In order to make the above finding quantitative, we consider the (inverse)
distribution function $X(F,T_1,T_2,\xi,r)$. We start by fixing $(T_1,T_2)$,
$\xi$ and some small probability $F$, which leaves us with a function of only
$r$. In order to obtain smoother interpolations for small
radius, however, we have used $N^{1/3}_r$ instead of $r$ 
as our independent variable, a technical detailed discussion can be found in Supplementary Note 3.

Fig.~\ref{fig:Xvsr_examples} shows plots of $1-X$
under these conditions, which exhibit well-defined 
peaks (see further information about the fitting function to the peaks in the Supplementary Note 2). Now, to a first approximation we can characterize 
any peak by its position, height and width. Fortunately, these three parameters turn
out to describe the scaling with $\xi$ of the full $1-X$
curve, see Supplementary Note 4.

The physical interpretation of the peak's parameters is clear. The peak's
height represents the strength of dynamic TC (the taller the peak, the larger
the chaos). The peak's position indicates the optimal lengthscale for the
study of TC, given the probability $F$, $\xi$ and the temperatures
$T_1,T_2$. The peak's width indicates how critical it is to spot this optimal
lengthscale (the wider the peak, the less critical the choice). Perhaps
unsurprisingly, the peak's position is found to scale linearly with $\xi$,
while the peak's width scales as $\xi^\beta$, with $\beta$ slightly larger than
one, see Supplementary Note 5 for further details. We shall focus here on the temperature and $\xi$
dependence of the peak's height (i.e., the strength of chaos), which has a
richer behavior.

\begin{figure}[t]
  \centering
  \includegraphics[width=.49\textwidth]{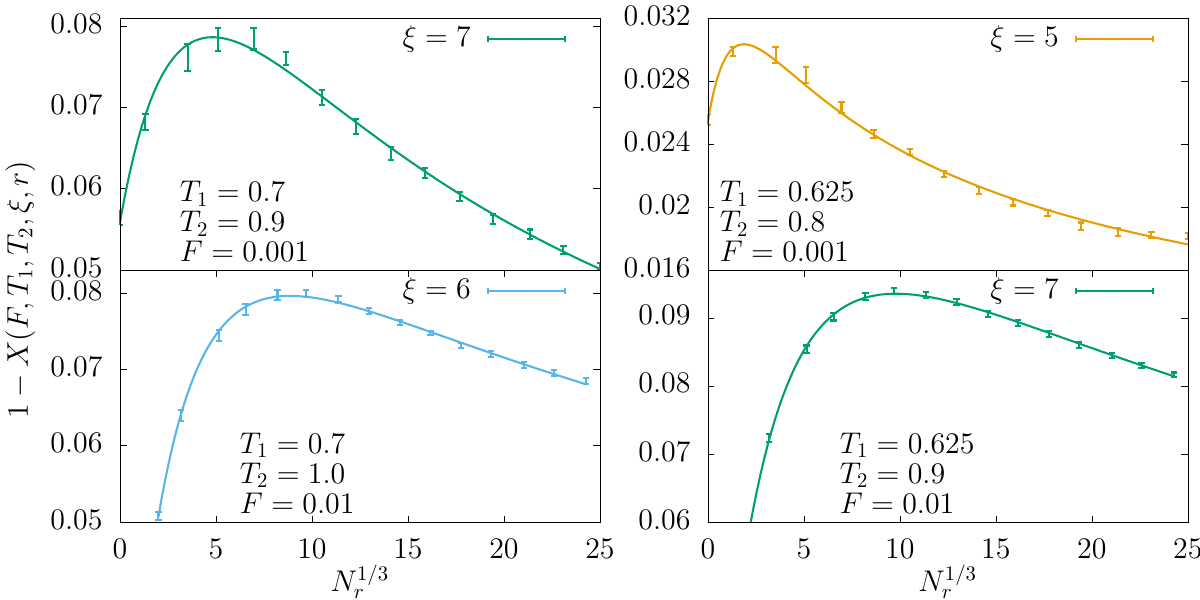}
  \caption{\textbf{Emergence of an optimal scale to observe temperature chaos.} The difference $1-X(F,T_1,T_2,\xi,r)$ [recall that
    $X(F,T_1,T_2,\xi,r)$ is the inverse of the distribution function] as a
function of the cubic root $N_r^{1/3}$ of the number of spins in the spheres, as
computed for different values of the probability level $F$, the temperatures $T_1$ and $T_2$, and the coherence length $\xi$. In this
representation, the optimal size of the spheres for the observation of
chaos (for given parameters $F$, $T_1$, $T_2$ and $\xi$) appears as the maximum
of the curves. Continuous lines are fits to a smooth 
interpolating function, further details can be found in Supplementary Note 2. Error bars represent one standard deviation.}
 \label{fig:Xvsr_examples}
\end{figure}

\begin{table}[t]
\begin{ruledtabular}
\begin{tabular}{c c c c c c c c c c c c c}
$F$ & \hspace{0.10cm} & $T_1$ &  \hspace{0.25cm} & $T_2$ & \hspace{0.10cm} & $\xi_{\min}$ & \hspace{0.10cm} & $\xi^*$ & \hspace{0.10cm} &$\alpha$ & \hspace{0.10cm} & $\chi^2/\mathrm{d.o.f}$ \\
\hline
0.001 & \hspace{0.10cm} & 0.625 &  \hspace{0.10cm} & 0.7 & \hspace{0.10cm} & 4.75 & \hspace{0.10cm} & 55(4) & \hspace{0.10cm} & 2.10(7) & \hspace{0.10cm} & 14.10/19 \\
0.001 & \hspace{0.10cm} & 0.625 &  \hspace{0.10cm} & 0.8 & \hspace{0.10cm} & 5.25 &\hspace{0.10cm} & 24.1(8) & \hspace{0.10cm} & 2.18(6) & \hspace{0.10cm} & 22.67/17\\
0.001 & \hspace{0.10cm} & 0.625 &  \hspace{0.10cm} & 0.9 & \hspace{0.10cm} & 4.75 &\hspace{0.10cm} & 16.8(3) & \hspace{0.10cm} & 2.09(4) & \hspace{0.10cm} & 28.88/19 \\
0.001 & \hspace{0.10cm} & 0.625 &  \hspace{0.10cm} & 1.0 & \hspace{0.10cm} & 4.75 &\hspace{0.10cm} & 13.24(15) & \hspace{0.10cm} & 2.04(3) & \hspace{0.10cm} & 8.77/19 \\
\hline 
0.001 & \hspace{0.10cm} & 0.7 &  \hspace{0.10cm} & 0.8 & \hspace{0.10cm} & 4.75 &\hspace{0.10cm} & 43.5(15) & \hspace{0.10cm} & 2.12(5) & \hspace{0.10cm} & 41.05/28 \\
0.001 & \hspace{0.10cm} & 0.7 &  \hspace{0.10cm} & 0.9 & \hspace{0.10cm} & 4.75 &\hspace{0.10cm} & 22.9(5) & \hspace{0.10cm} & 2.09(4) & \hspace{0.10cm} & 33.32/28 \\
0.001 & \hspace{0.10cm} & 0.7 &  \hspace{0.10cm} & 1.0 & \hspace{0.10cm} & 4.75 &\hspace{0.10cm} & 16.3(2) & \hspace{0.10cm} & 2.04(4) & \hspace{0.10cm} & 22.32/28 \\
\hline
0.01 & \hspace{0.10cm} & 0.625 &  \hspace{0.10cm} & 0.8 & \hspace{0.10cm} & 5.75 &\hspace{0.10cm} & 29.3(5) & \hspace{0.10cm} & 2.21(3) & \hspace{0.10cm} & 13.32/15 \\
0.01 & \hspace{0.10cm} & 0.625 &  \hspace{0.10cm} & 0.9 & \hspace{0.10cm} & 5.75 &\hspace{0.10cm} & 20.5(3) & \hspace{0.10cm} & 2.12(2) & \hspace{0.10cm} & 16.05/15\\
0.01 & \hspace{0.10cm} & 0.625 &  \hspace{0.10cm} & 1.0 & \hspace{0.10cm} & 4.75 &\hspace{0.10cm} & 15.87(16) & \hspace{0.10cm} & 2.08(2) & \hspace{0.10cm} & 23.93/19 \\
\hline 
0.01 & \hspace{0.10cm} & 0.7 &  \hspace{0.10cm} & 0.8 & \hspace{0.10cm} & 4.75 &\hspace{0.10cm} & 51.4(12) & \hspace{0.10cm} & 2.17(3) & \hspace{0.10cm} & 8.06/28 \\
0.01 & \hspace{0.10cm} & 0.7 &  \hspace{0.10cm} & 0.9 & \hspace{0.10cm} & 5.25 &\hspace{0.10cm} & 27.9(4) & \hspace{0.10cm} & 2.11(2) & \hspace{0.10cm} & 31.56/26 \\
0.01 & \hspace{0.10cm} & 0.7 &  \hspace{0.10cm} & 1.0 & \hspace{0.10cm} & 4.75 &\hspace{0.10cm} & 19.9(2) & \hspace{0.10cm} & 2.05(2) & \hspace{0.10cm} & 31.78/28 \\
\end{tabular}
\end{ruledtabular}
\caption{\textbf{Parameters describing the crossover between weak and strong temperature chaos regimes.} Parameters obtained in the fits to Eq.~\eqref{eq:fmax_xi} of our data
  for the peak's height, see Fig.~\ref{fig:Xvsr_examples}, with
  $\xi_{\text{min}}\leq \xi \leq \xi_{\text{max}}$. We also report the fits'
  figure of merit $\chi^2/\mathrm{d.o.f}$. We chose $\xi_{\text{min}}$ by
  requiring a $P$ value greater than $0.05$ in the fits ($\xi_{\text{max}}=9.5$
  for $T_1=0.625$ and $\xi_{\text{max}}=12.5$ for $T_1=0.7$). $T_1$ and $T_2$ represent the temperatures involved in the computation of the chaotic parameter. Unfortunately,
  the flatness of the peak for ($T_1=0.625,T_2=0.7,F=0.01$) did not allow us
  to compute the peak's parameters.}
\label{tab:parametros_fmax}
\end{table}

The $\xi$ dependence of the
peak's height  (for a given probability $F$ and temperatures $T_1$ and $T_2$)
turns out to be reasonably well described by the following
ansatz:
\begin{equation}\label{eq:fmax_xi}
  f_{\max}(\xi) = \dfrac{\varepsilon(\xi)}{1+ \varepsilon(\xi)} \, ,
\text{ with } \varepsilon(\xi)=(\xi/\xi^*)^\alpha\,.
\end{equation}
This formula describes a crossover phenomenon, ruled by a characteristic length
$\xi^*$. For $\xi\ll\xi^*$ the peak's height grows with $\xi$ as a power law,
while for $\xi\gg\xi^*$ the strong-chaos limit [i.e.,  $(1-X)\to 1$ ] is
approached. However, some consistency requirements should be met before taking
the crossover length $\xi^*$ seriously. Not only should the fit to
Eq.~\eqref{eq:fmax_xi} be of acceptable statistical quality (the fit
parameters are the characteristic lengthscale $\xi^*$ and the exponent
$\alpha$). One would also wish exponent $\alpha$ to be independent of the
temperatures $T_1$ and $T_2$  and of the chosen probability $F$.

We find fair fits to Eq.~\eqref{eq:fmax_xi}, see
Table~\ref{tab:parametros_fmax}. In all cases, exponent $\alpha$ turns out to
be compatible with $2.1$ at the two-$\sigma$ level [except for the
$(F=0.01,T_1=0.625,T_2=0.8)$ fit]. Under these conditions, we can interpret
$\xi^*$ as a characteristic length indicating the crossover from weak to
strong TC, at the probability level indicated by $F$. Furthermore, the
relatively large value of exponent $\alpha$ indicates that this crossover is
sharp.

\begin{figure}[t]
	\centering
	\includegraphics[width=0.49\textwidth]{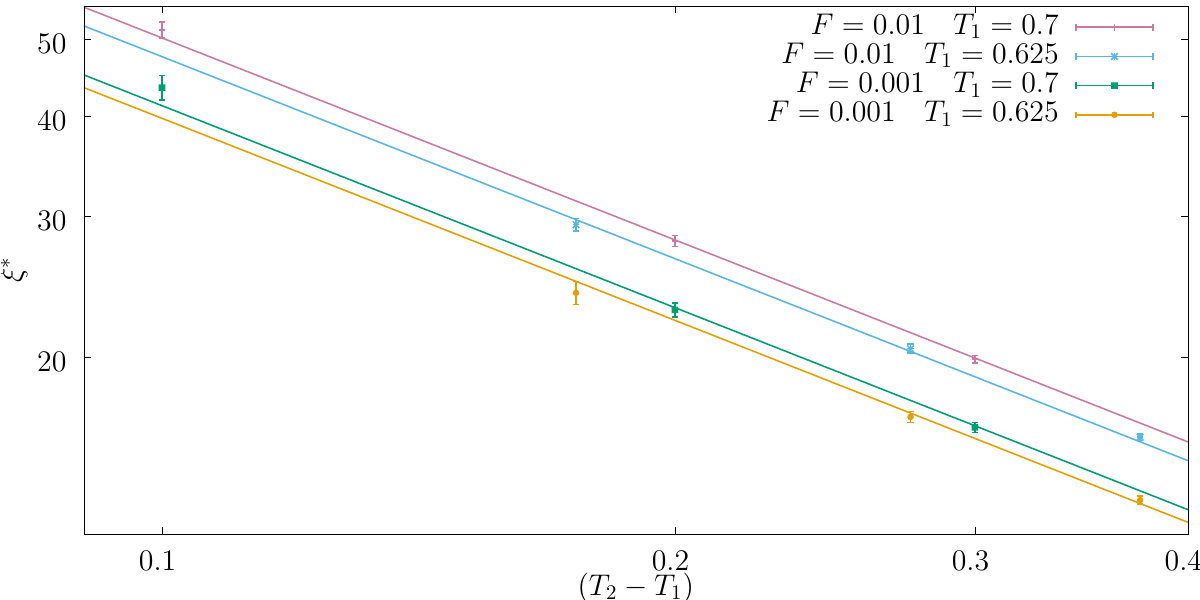}
	\caption{\textbf{Universal scaling of dynamic chaos.} The characteristic length $\xi^*$ is
plotted against the temperature difference $T_2-T_1$ in a log-log scale. Each
curve is uniquely identified by the probability level $F$ and the smallest
temperature of each pair $T_1$. Fits to Eq.~\eqref{eq:def_zeta}, enforcing a
common exponent, are shown with continuous lines and result in a chaotic
exponent $\zeta_\text{NE}=1.19(2)$. Error bars represent one standard deviation.}
 \label{fig:zeta_exponent}
\end{figure}

The trends for the crossover length $\xi^*$ in Table~\ref{tab:parametros_fmax}
are very clear: $\xi^*$ grows upon increasing $F$ or upon decreasing
$T_2-T_1$. Identifying $\xi^*$ as the non-equilibrium partner of the
equilibrium chaotic length $\ell_\text{c}(T_1,T_2)$~\cite{fisher:86,bray:87b}
will allow us to be more quantitative (indeed, the two lengthscales
indicate the crossover between weak chaos and strong chaos). Now, the
equilibrium $\ell_\text{c}(T_1,T_2)$ has been found to scale for the 3D Ising
spin glass as
\begin{equation}
\ell_\text{c}(T_1,T_2) \propto (T_2-T_1)^{-1/\zeta} \, \, , \label{eq:def_zeta_equilibrium}
\end{equation}
with $\zeta \approx 1.07$~\cite{katzgraber:07} or
$\zeta \approx 1.07(5)$~\cite{fernandez:13}. These considerations suggest the
following ansatz for the non-equilibrium crossover length
\begin{equation}
\xi^*(T_1,T_2,F) = B(F,T_1) \, (T_2-T_1)^{-1/\zeta_{\text{NE}}} \, \, , \label{eq:def_zeta}
\end{equation}
where $B(F,T_1)$ is an amplitude. We have tested Eq.~\eqref{eq:def_zeta} 
by computing a joint fit for four $(T_1, F)$ pairs as functions of $T_2-T_1$, allowing
each curve to have its own amplitude but enforcing a common $\zeta_\text{NE}$ (see Fig.~\ref{fig:zeta_exponent}).
The resulting $\chi^2/\text{d.o.f.} = 7.55/7$ validates our ansatz, with
an exponent $\zeta_\text{NE}=1.19(2)$ fairly close 
to the equilibrium result $\zeta=1.07(5)$~\cite{fernandez:13}. This agreement
strongly supports our physical interpretation of the crossover
length. We, furthermore, find that $B$ is only weakly dependent on $T_1$.
Nevertheless, the reader should be warned that it has been
suggested~\cite{fernandez:13} that the equilibrium exponent $\zeta$ may be
different in the weak- and strong-chaos regimes.

\section*{Conclusions}
 We have shown that the concept of temperature chaos can
be meaningfully extended to the non-equilibrium dynamics of a large spin
glass. This is, precisely, the framework for rejuvenation and memory
experiments~\cite{jonason:98,lundgren:83,jonsson:00,hammann:00}, as well as
other more chaos-oriented experimental work~\cite{zhai:20b}. Therefore, our
precise characterization of dynamical temperature chaos paves the way for the
interpretation of these and forthcoming experiments.  Our simulation of spin-glass 
dynamics doubles the numerical effort in~\cite{janus:18} and has been
carried out on the Janus-II special-purpose supercomputer.

The key quantity governing dynamic temperature chaos is the time-dependent
spin-glass coherence length $\xi(\tw)$. The very strong spatial
heterogeneity of this phenomenon is quantified through a 
distribution function $F$. This probability can be
thought of as the fraction of the sample that shows a chaotic response to
a given degree. When comparing temperatures $T_1$ and $T_2$, the degree of
chaoticity is governed by a lengthscale $\xi^*(F,T_1,T_2)$. While chaos is
very weak if $\xi(\tw)\ll\xi^*(F,T_1,T_2)$, it quickly becomes strong as
$\xi(\tw)$ approaches $\xi^*(F,T_1,T_2)$. We find that, when $T_1$ approaches
$T_2$, $\xi^*(F,T_1,T_2)$ appears to diverge with the same critical exponent
that it is found for the equilibrium chaotic length~\cite{fernandez:13}.

Although we have considered in this work fairly small values of the chaotic
system fraction $F$, a simple extrapolation, linear in $\log F$, predicts
$\xi^* \approx 60$ for $F=0.1$ at $T_1=0.7$ and $T_2=0.8$ (our closest pair
of temperatures in Table~\ref{tab:parametros_fmax}). A spin-glass coherence
length well above $60 a_0$ is experimentally reachable
nowadays~\cite{zhai:20b,zhai-janus:20,zhai-janus:21,zhai:19} ($a_0$ is the typical spacing
between spins), which makes our dynamic temperature chaos significant. Indeed,
while completing this manuscript, a closely related experimental
study~\cite{zhai:20b} reported a value for exponent $\zeta_{\text{NE}}$ in
fairly good agreement with our result of $\zeta_\text{NE}=1.19(2)$ 
in Fig.~\ref{fig:zeta_exponent}.

Let us conclude by commenting on possible venues for future
  research. Clearly, it will be important to understand in detail how dynamic
  temperature chaos manifests itself in non-equilibrium experiments. Simple
  protocols (in which temperature sharply drops from $T_2$ to $T_1$, see,
  e.g., Zhai et. al.\cite{zhai:20b}) seems more accessible to a first analysis than
  memory and rejuvenation
  experiments~\cite{jonason:98,lundgren:83,jonsson:00,hammann:00}. An
  important problem is that the correlation functions that are studied
  theoretically are not easily probed experimentally. Instead,
  experimentalists privilege the magnetization density (which is a spatial
  average over the whole sample). Therefore an important theoretical goal is to
  predict the behavior of the non-equilibrium time-dependent magnetization
  upon a temperature drop. One may speculate that the Generalized
  Fluctuation-Dissipation Relations~\cite{cugliandolo:93} might be the route
  connecting the correlation functions with the response to an externally
  applied magnetic field.  Interestingly enough, these relations (that apply
  at fixed temperature) can be defined locally as well~\cite{castillo:02}. The
  resulting spatial distribution function allows the reconstruction of the
  global response to the magnetic field. Extending this analysis to a
  temperature drop may turn out to be fruitful in the future.

\section*{Methods} 
All the observables involved in the computation of temperature chaos depend
on a pair of replicas $(\sigma,\tau)$. The basic quantity is the overlap field
\begin{equation} q^{\sigma,\tau}(\mathbf{x},\tw) = s_{\mathbf{x}}^\sigma(\tw) s^\tau_{\mathbf{x}}(\tw) \>\> ,
\end{equation}
 Usually, this pair of replicas are at
the same temperature $T$. All the definitions are, however,
straightforwardly extended to two temperatures.  For instance, the four-point
two-temperature spatial correlation function is
\begin{widetext}
\begin{equation}
C_4^{T_1T_2}(T_1,T_2,t_{\mathrm{w}1},t_{\mathrm{w}2},\mathbf{r}) = \left[\langle q^{\sigma(T_1),\tau(T_2)} (\mathbf{x},t_{\mathrm{w}1},t_{\mathrm{w}2})
q^{\sigma(T_1),\tau(T_2)}(\mathbf{x}+\mathbf{r},t_{\mathrm{w}1},t_{\mathrm{w}2})\rangle_{T}\right]_J \>\>, \label{eq:def_corr_func_2T}
\end{equation}
\end{widetext}
where $[\ldots]_J$ denotes the average over the samples.
Building on this function we can define our integral estimator for the coherence
length~\cite{janus:09b}:
\begin{equation} I^{T_1T_2}_k( t_{\mathrm{w}1},t_{\mathrm{w}2}) = \int_{0}^{\infty}r^k\,C^{T_1T_2}_4(r,t_{\mathrm{w}1},t_{\mathrm{w}2})\,\mathrm{d} r \>\>
, \label{eq:def_integral_2T}
\end{equation}
and
\begin{equation} \xi^{T_1T_2}_{k,k+1}(t_{\mathrm{w}1},t_{\mathrm{w}2}) = \dfrac{I^{T_1T_2}_{k+1}(t_{\mathrm{w}1},t_{\mathrm{w}2})}{I^{T_1T_2}_k(t_{\mathrm{w}1},t_{\mathrm{w}2})}
\>\> . \label{eq:def_correlation_length_2T}
\end{equation}
As explained in the main text, times $t_{\mathrm{w}1}$ and
$t_{\mathrm{w}2}$ are fixed through the condition expressed in
Eq.~\eqref{eq:el_reloj_doble}, which ensures that we are comparing
spin configurations that are ordered on the same length scale.

Since our $\tw$ are on a discrete grid, we solve
Eq.~\eqref{eq:el_reloj_doble} for the global overlaps through a
(bi)linear interpolation.

\section*{Data Availability}
The data contained in the figures of this paper, accompanied by the gnuplot
script files that generate these figures, are publicly available at
\href{https://github.com/JanusCollaboration/caosdin}{https://github.com/JanusCollaboration/caosdin}.

The data that support the findings of this study are available from the corresponding author upon reasonable request.

\section*{Code availability}
The codes that support the findings of this study are available from the corresponding author upon reasonable request.

\begin{acknowledgments}
  We are grateful for discussions with R. Orbach and Q. Zhai.  This work was
  partially supported by Ministerio de Econom\'ia, Industria y Competitividad
  (MINECO, Spain), Agencia Estatal de Investigaci\'on (AEI, Spain), and Fondo
  Europeo de Desarrollo Regional (FEDER, EU) through Grants
  No. FIS2016-76359-P, No.  PID2019-103939RB-I00, No.  PGC2018-094684-B-C21
  and PGC2018-094684-B-C22, by the Junta de Extremadura (Spain) and Fondo
  Europeo de Desarrollo Regional (FEDER, EU) through Grant No.  GRU18079 and
  IB15013 and by the DGA-FSE (Diputaci\'on General de Arag\'on -- Fondo Social
  Europeo).  This project has also received funding from the European Research
  Council (ERC) under the European Union's Horizon 2020 research and
  innovation program (Grant No. 694925-LotglasSy). DY was supported by the
  Chan Zuckerberg Biohub and IGAP was supported by the Ministerio de Ciencia,
  Innovaci\'on y Universidades (MCIU, Spain) through FPU grant No. FPU18/02665.
  BS was supported by the Comunidad de Madrid and the Complutense University
  of Madrid (Spain) through the Atracci\'on de Talento program (Ref.
  2019-T1/TIC-12776).
\end{acknowledgments}

\section*{Author Contributions}
J.M.~G-N and D.~N contributed to Janus/Janus II simulation software. D.~I, A.~T and R.~T contributed to Janus II design. M.~B-J, E.~C, A.~C, L.A.~F, J.M.~G-N, I.~G-A~P, A.~G-G, D.~I, A.~M, A.~M-S, I.~P,S.~P-G, S.F.~S, A.~T and R.~T contributed to Janus II hardware and software development. L.A.~F, V.~M-M and J.~M-G designed the research. J.~M-G analyzed the data. M.~B-J, L.A.~F, E.~M,V.~M-M, J.~M-G, I.~P, G.~P, B.~S, J.J.~R-L, F.~R-T and D.~Y discussed the results. V.~M-M, J.~M-G, B.~S and D.~Y wrote the paper.

\section*{Supplementary Note 1: Extrapolation to infinite replicas}
\begin{figure}[!h]
  \centering
  \includegraphics[width=0.49\textwidth]{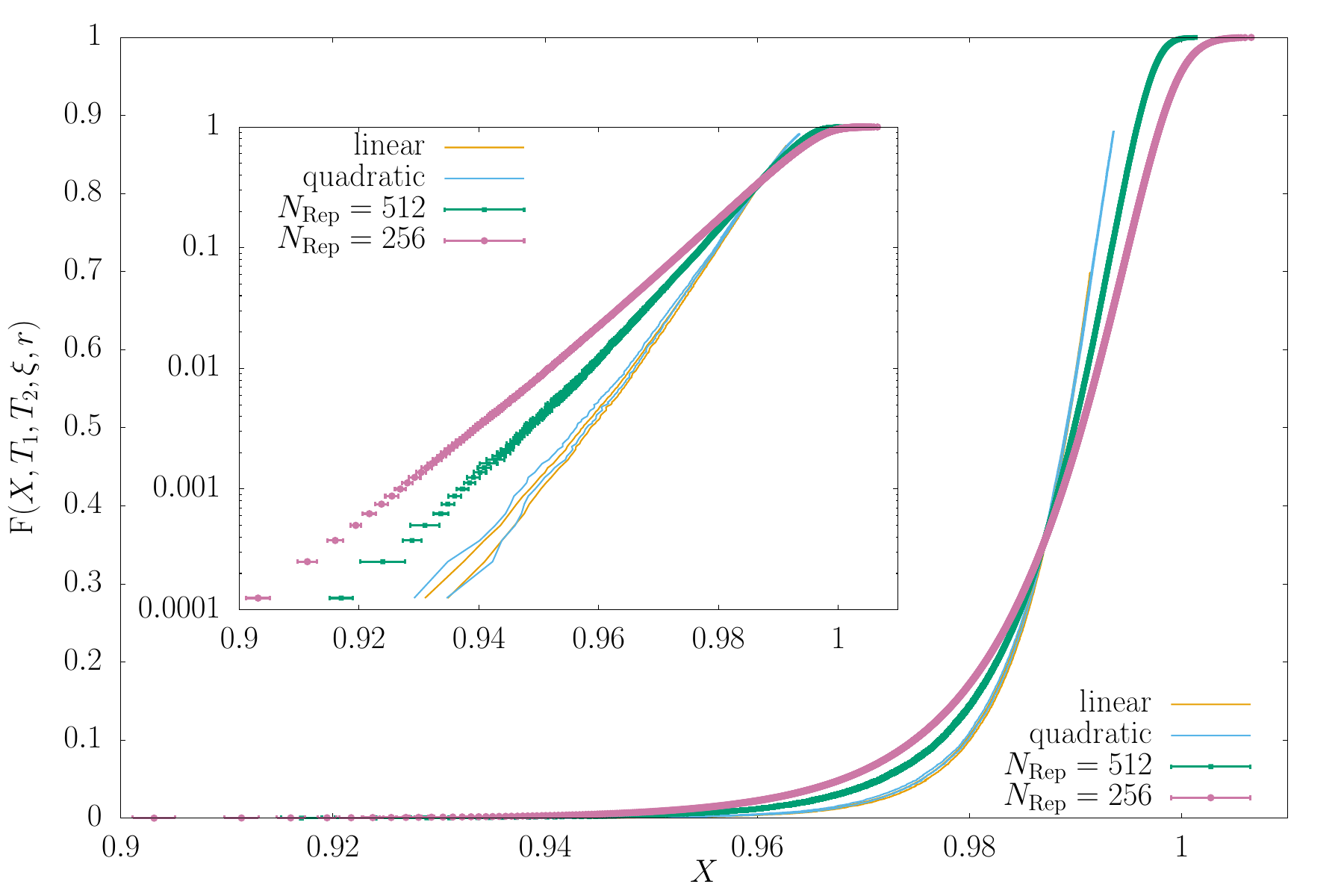}
  \caption{Linear and quadratic extrapolations,
Eq.~\eqref{eq:extrapolacion_lineal} and
Eq.~\eqref{eq:extrapolacion_cuadratica}, turn out to be equivalent for the tail
of the distribution function. Continuous lines are the linear (golden
curves) and quadratic (blue curves) extrapolations to $\NRep \to \infty$ for
$F(X,T_1,T_2,\xi,r)$ as a function of $X$. The data shown correspond to the
case $T_1=0.7$, $T_2=0.8$, $\xi=11$ and $r=8$. The two curves shown for each
extrapolation correspond to the central value plus or minus the standard error.
We show horizontal error bars because we are computing the inverse distribution
function $X(F,T_1,T_2,\xi,r)$. We only show extrapolated data when
$\chi^2/\text{d.o.f} < 1.1$ in the fits to Eq.~\ref{eq:extrapolacion_lineal} or
to Eq.~\ref{eq:extrapolacion_cuadratica}. For comparison, we also plot the data
corresponding to $\NRep=512$ and $\NRep=256$. \textit{Inset:} As in the main plot, but with the vertical axis
in log scale.}
\label{fig:linear_quadratic}
\end{figure}

\begin{figure}[!h]
  \centering
  \includegraphics[width=0.49\textwidth]{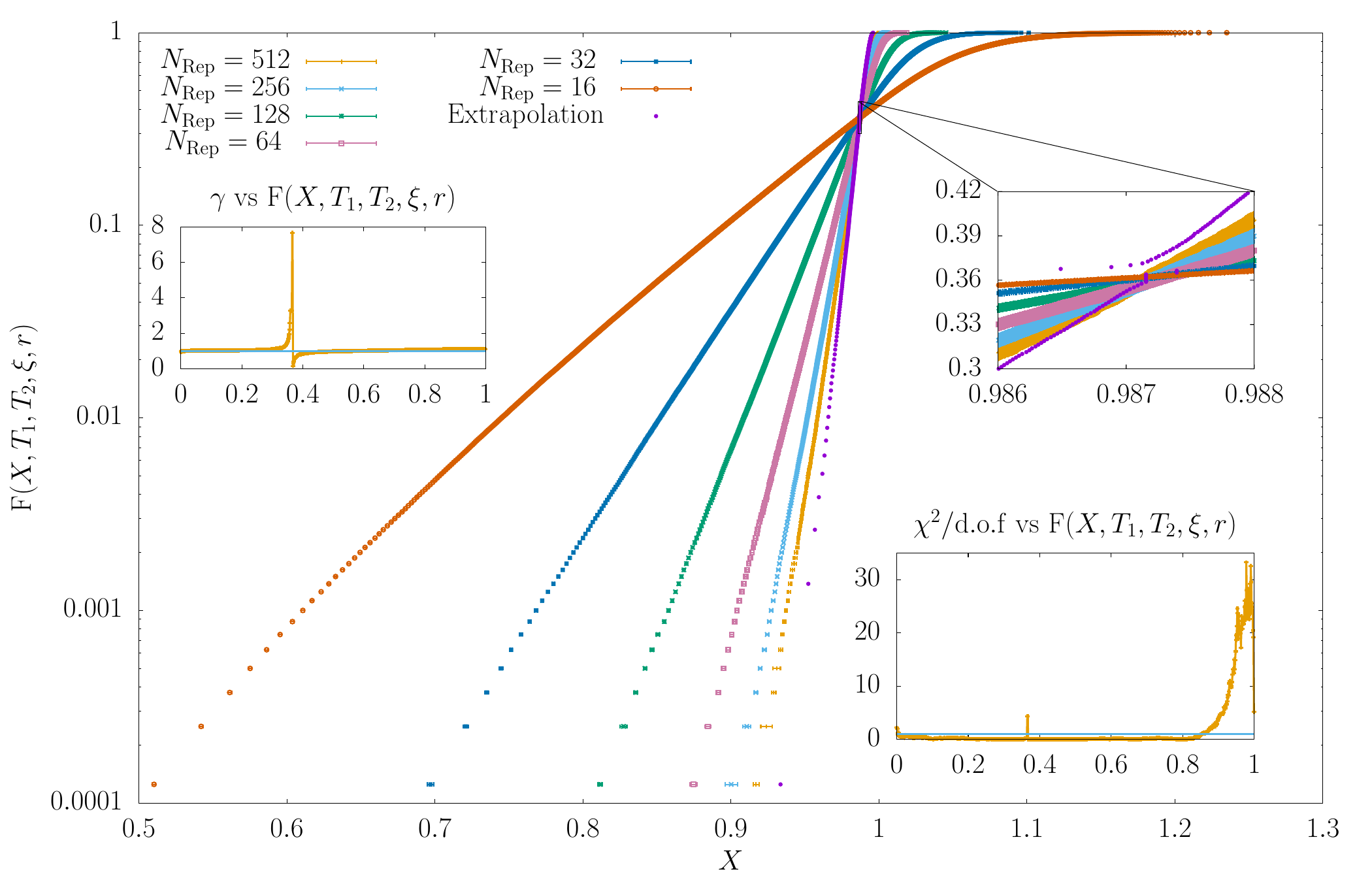}
  \caption{The exponent $\gamma$ in Eq.~\eqref{eq:extrapolacion_libre},
remains close to one when it becomes a fit parameter. The distribution
function $F(X,T_1,T_2,\xi,r)$ is plotted as a function of $X$ for $\NRep=\{512,
256, 128 ,64, 32, 16\}$ together with the extrapolation to $\NRep \to \infty$,
as obtained from a fit to Eq.~\eqref{eq:extrapolacion_libre}. The data shown
correspond to $T_1=0.7$, $T_2=0.8$, $\xi=11$ and $r=8$. In order not to clutter
the figure, we do not show error bars, that represent one standard deviation, in the $\NRep \to \infty$ extrapolation.
\textit{Left inset:} Exponent $\gamma$ plotted against the
probability $F$. The exponent remains close to $\gamma=1$ for all $F$, with the exception of
the unstable behavior at $F \approx 0.35$, where curves for different $\NRep$
cross (see also \textit{top-right inset}). \textit{Bottom-right inset:}
Goodness-of-fit estimator $\chi^2$ per
degree of freedom plotted against $F$. The blue line corresponds to
$\chi^2/\text{d.o.f} = 1$.  \textit{Top-right inset:} Closeup of the main plot,
emphasizing the crossing region at $F \approx 0.35$. Note that at that particular
value of $F$ the data shows almost no dependence with $\NRep$, which makes 
the fit to Eq.~\eqref{eq:extrapolacion_libre} unstable.} \label{fig:exponente_libre}
\end{figure}

The thermal expectation values necessary to compute the
chaotic parameter are defined in the limit of infinite replicas,
so an extrapolation is necessary to avoid bias.
Fortunately, with all other parameters fixed, the evolution
 of $X(F,T_1,T_2,\xi,r)$ with $\NRep$ is smooth (see Fig.~\ref{fig:linear_quadratic} and
Fig.~\ref{fig:exponente_libre}). We have mainly used a linear
ansatz for the extrapolation,
\begin{equation}
X_{\NRep} = X_{\infty} + \dfrac{A}{\NRep} \>\>\> , \label{eq:extrapolacion_lineal}
\end{equation}
where  $X_{\NRep}$ is shorthand for
$X(F,T_1,T_2,\xi,r;\NRep)$,
$X_\infty=X(F,T_1,T_2,\xi,r; \NRep\to\infty)$
and $A$ is a constant. As a check for the linear ansatz
in Eq.~\eqref{eq:extrapolacion_lineal}, we have considered two alternative
functional forms:
\begin{align}
X_{\NRep} &= X_{\infty} + \dfrac{B}{\NRep} + \dfrac{C}{\NRep^2} \>\>\> , \label{eq:extrapolacion_cuadratica}\\
X_{\NRep} &= X_{\infty} + \dfrac{D}{\NRep^\gamma} \>\>\> , \label{eq:extrapolacion_libre}
\end{align}
where $B$, $C$ and $D$ are amplitudes and $\gamma$ is a free exponent. We
perform independent fits to Eq.~\eqref{eq:extrapolacion_lineal},
Eq.~\eqref{eq:extrapolacion_cuadratica} and Eq.~\eqref{eq:extrapolacion_libre}
for every value of the parameters $(F,T_1,T_2,\xi,r)$. We reject fits with a
diagonal $\chi^2/\text{d.o.f}\geq 1.1$. Errors in $X_{\infty}$ are computed
by performing separate
fits for each jackknife block (the fitting procedure consists in minimizing
the diagonal $\chi^2$, see~\cite{janus:09b}).

As a first check, we compare the linear and quadratic extrapolations (see
Fig.\ref{fig:linear_quadratic} for an illustrative example). The figure shows
that even for our largest $\NRep$, namely $\NRep=256$ and $\NRep=512$, we are
still far from the $\NRep \to \infty$ limit. Fortunately,
the linear and the quadratic extrapolations yield compatible results in our
region of interest, i.e., the tail of the distribution function. We remark that
the consistency condition $\chi^2/\text{d.o.f} < 1.1$ is met in a larger range
for the quadratic extrapolation ($F<0.9$) than for the linear extrapolation
($F<0.7$). However, because both coincide in the low-$F$ range we are
interested in, we have kept the simpler linear extrapolation.

Our second check in Eq.~\eqref{eq:extrapolacion_libre} seeks the natural
exponent $\gamma$ for the extrapolation as a fitting parameter. We have found
that the consistency condition $\chi^2 / \mathrm{d.o.f}<1.1$ is met for
$F<0.85$. Fortunately, $\gamma$ turns out
to be very close to the value $\gamma=1$, with the exception of an instability
in the crossing region around $F\approx 0.35$, see Fig.~\ref{fig:exponente_libre}. 

In summary, the quadratic and the free-exponent extrapolations support our
choice of Eq.~\ref{eq:extrapolacion_lineal} as the preferred form for the
$\NRep \to \infty$ extrapolation.

\section*{Supplementary Note 2: Characterization of the peak} \label{sect:characterization}
The complementary of the chaotic parameter $1-X$, as a function of the sphere
size, has a well-defined peak. Characterizing the peak is fundamental
to the analysis because it provides information about the optimal sphere size
for the study of temperature chaos and about the strength of the phenomenon.

Let us remark that, at least close to a maximum, any smooth curve is
characterized by the position, height and width of the peak. In order
to meaningfully compute these three parameters from our data, we fit $1-X$ to
Eq.~\eqref{eq:functional_form} with $z=N_r^{1/3}$.
We extract the position, width and height of the maximum from the fitted
function $f(z)$. Errors are computed with a jackknife method~\cite{janus:09b}.

\section*{Supplementary Note 3: On the most convenient variable to characterize the sphere size}\label{sect:sphere_size}
\begin{figure}[!h]
  \centering
  \includegraphics[width=0.49\textwidth]{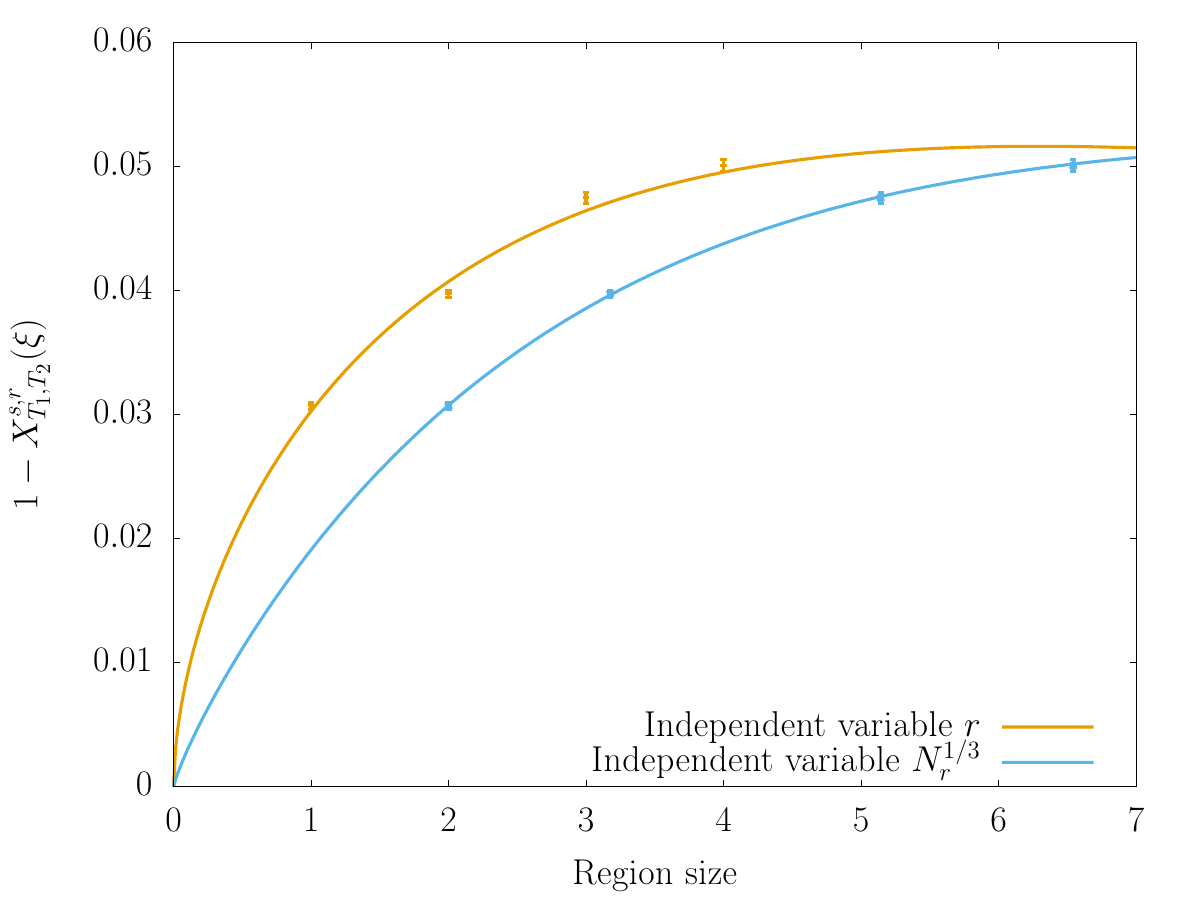}
  \includegraphics[width=0.47\textwidth]{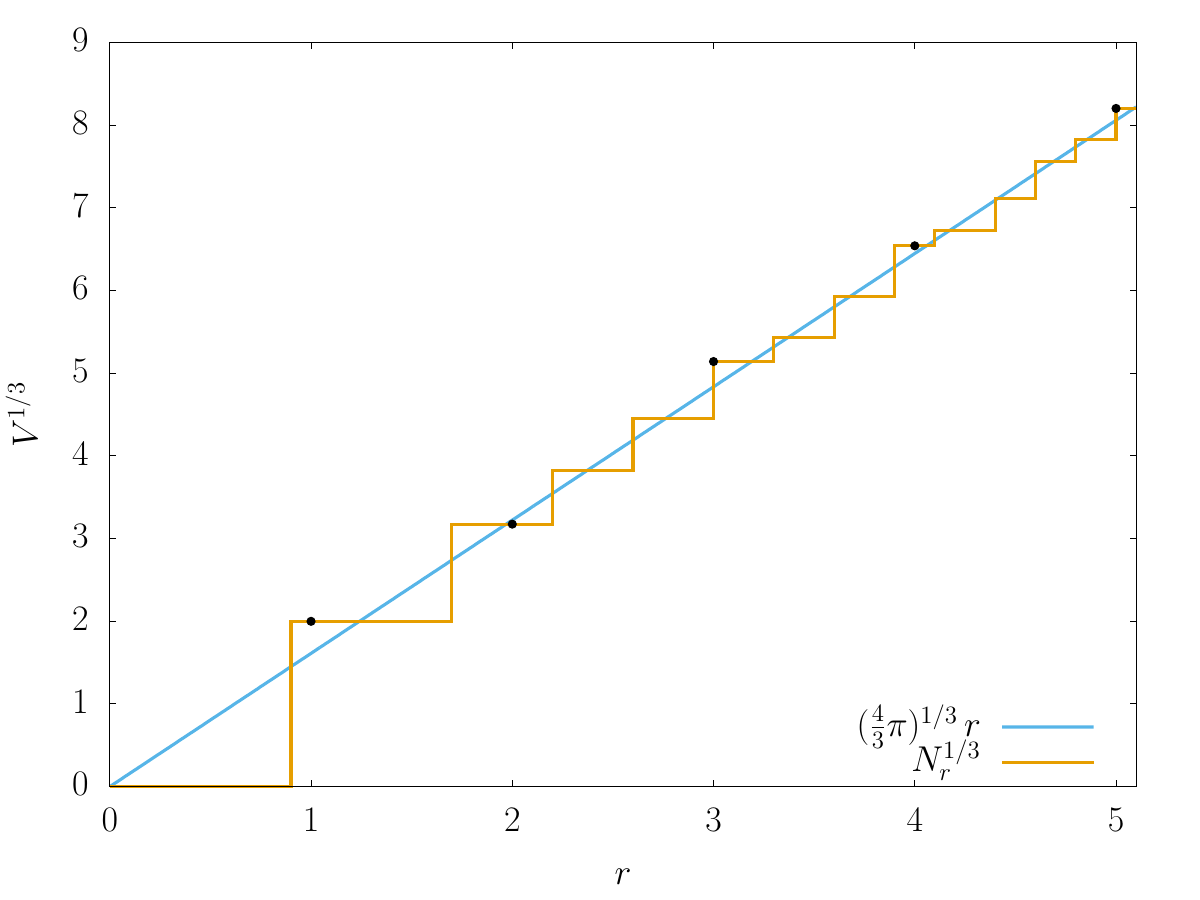}	
  \caption{$N_r^{1/3}$ is a better variable to describe
short length scales. \textit{Upper panel:} complementary of temperature chaos
$1-X^{s,r}_{T_1,T_2}(\xi)$ against the region size for the two discussed
independent variables, namely $N_r^{1/3}$ and the radius $r$. The continuous
lines are fits to Eq.~\eqref{eq:functional_form} taking as variable either $z=r$
(golden curve) or $z=N_r^{1/3}$ (blue curve). The shown data corresponds to
$T_1=0.7$, $T_2=0.9$, $F=0.01$ and $\xi=7$. We enlarge the small-sphere region,
 where both independent variables most differ. Error bars represent one standard deviation. \textit{Lower panel:}
the cubic root of the volume of a sphere (blue curve) is plotted as a function
of the radius of the sphere $r$. The golden curve is $N_r^{1/3}$, namely the
cubic root of the number of lattice points contained in a sphere of radius $r$,
centered at a node of the dual to our cubic lattice.
Values of $N_r^{1/3}$ corresponding to integer $r$ are highlighted as black
dots.}
 \label{fig:Nr_vs_r}
\end{figure}

In this section we explain our rationale for choosing the cubic root of the
number of spins contained in the sphere, $N_r^{1/3}$, rather than its radius
$r$, to characterize the size of the spheres considered in our analysis.

The objective is fitting the peaks of $1-X$ to a function of the form
\begin{equation}\label{eq:functional_form}
  f(z) = \dfrac{az^b}{1+cz^d} \, ,
\end{equation}
with $a$, $b$, $c$, and $d$ as fit parameters. If one uses
the obvious choice of $z=r$, however, the fit fails.  Indeed, see
Fig.~\ref{fig:Nr_vs_r} upper panel, $1-X$ is not a smooth function of $r$.
The reason is that the number of lattice points in the
spheres is not a smooth function of $r$ either (see Fig.~\ref{fig:Nr_vs_r}
lower panel). It is natural, therefore, to replace $r$ with $N_r^{1/3}$
as independent variable. This substitution makes Eq.~\eqref{eq:functional_form}
work down to smaller spheres.  The difference between both
independent variables becomes negligible for very large spheres.

\section*{Supplementary Note 4: Global versus local description of the peaks}
In the main text, we reduce the study of the scaling of temperature chaos with
the coherence length $\xi$ to the study of the peak of $1-X$ against the size
of the spheres. The reader may wonder whether the local fit
of the peak will extend to describe the  full curve. Here we present some
positive evidences in this respect.

\begin{figure}[!h]

  \centering
  \includegraphics[width=0.45\textwidth]{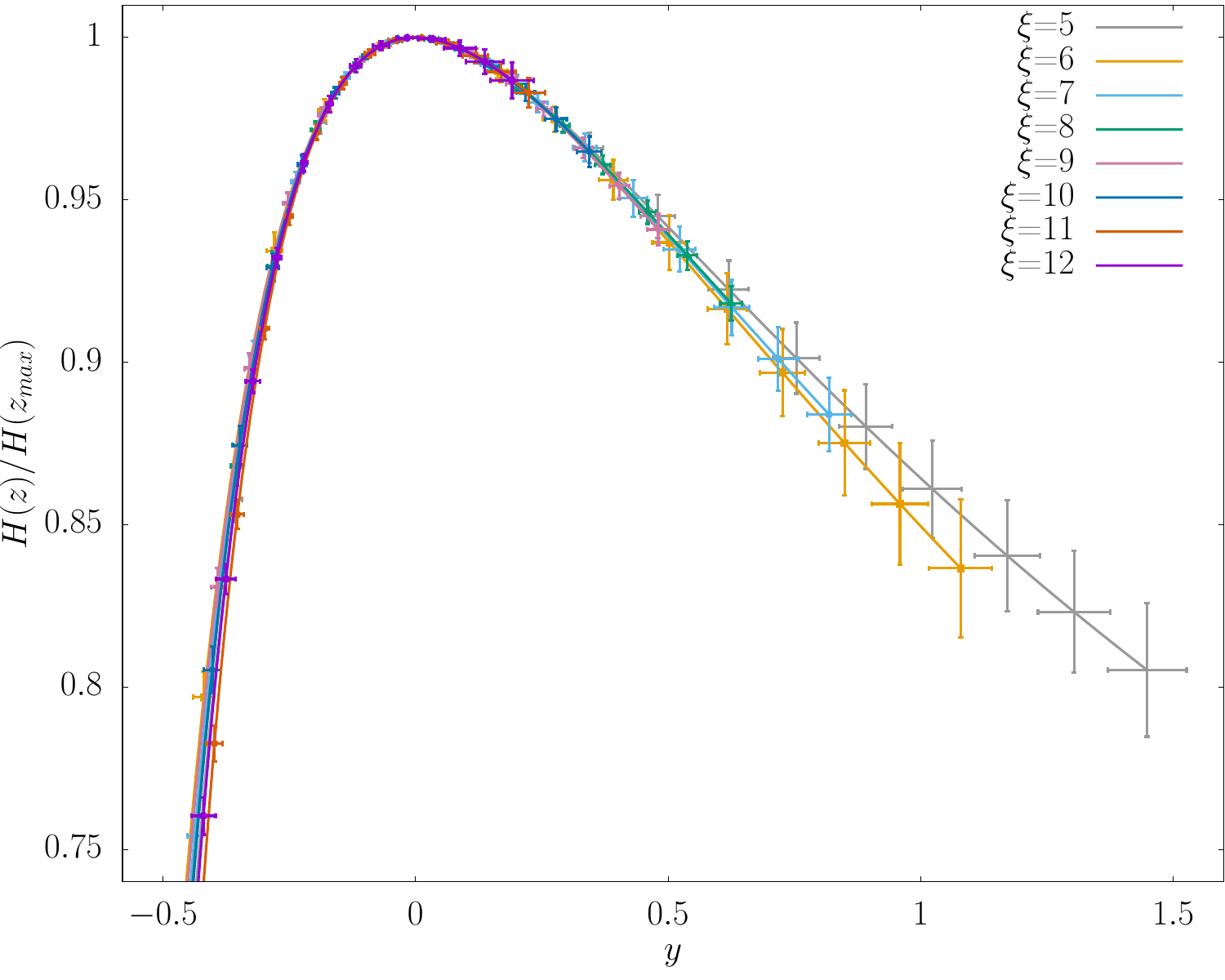}
  \includegraphics[width=0.45\textwidth]{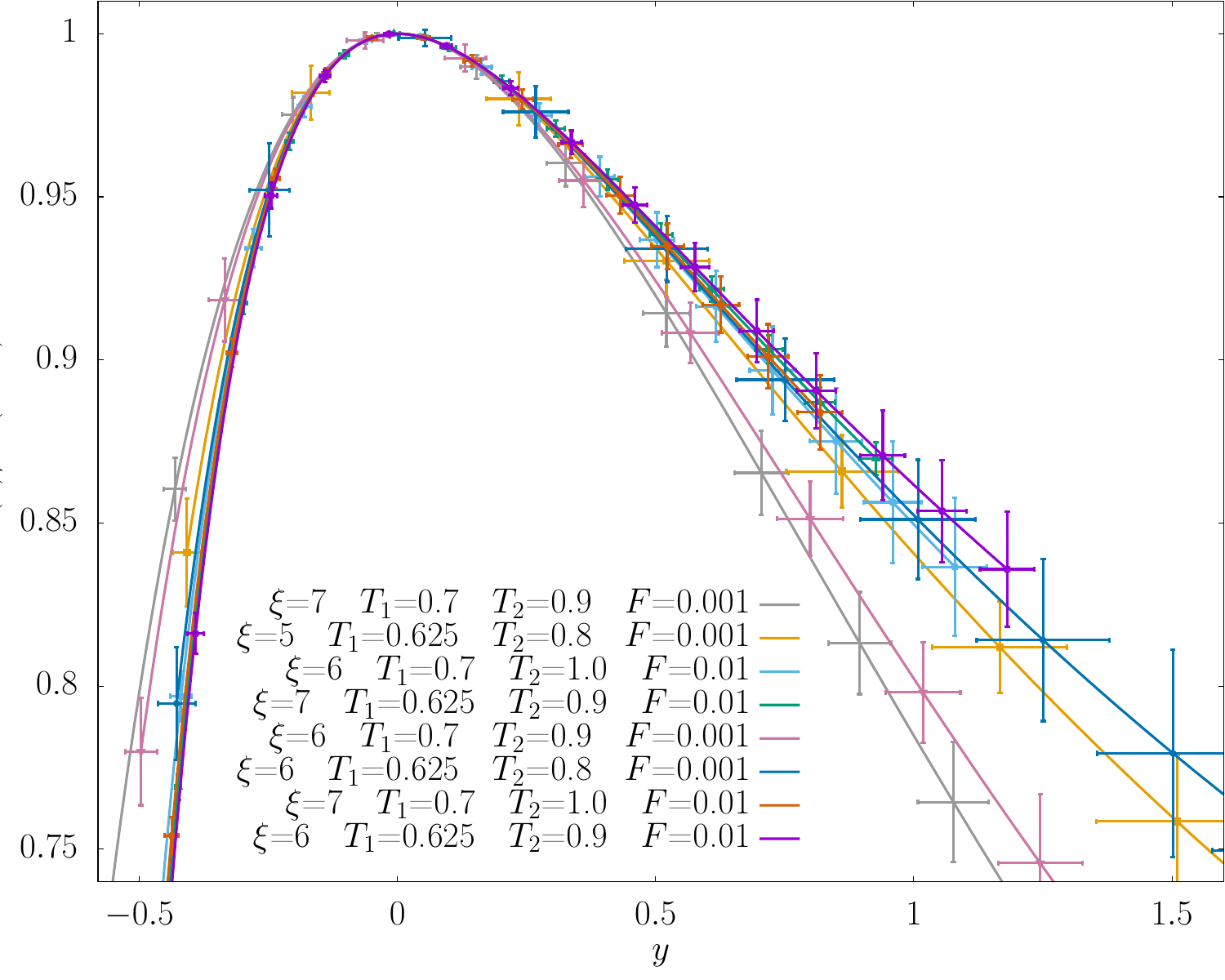}
  \caption{{Universality in $1-X$ extends beyond the trivial Taylor's
universality.} The upper panel shows $1-X$ in units of its peak value, for the
temperatures $T_1=0.7$, $T_2=1.0$ and $F=0.01$. Taylor's theorem implies that,
using the independent variable $y$ [see Eq.~\eqref{eq:Taylor-universality}],
the different curves should coincide close to $y=0$. However, we see that the
collapse holds beyond the quadratic approximation (as evinced by the strong
asymmetry of the master curve). The lower panel mixes different values of $F$,
$T_1$ and $T_2$, which leads to significant discrepancies for large values of
$|y|$. Nevertheless, the curves still collapse in a range
$y \in (-0.3,0.5)$ where the asymmetry is significative. Error bars represent one standard deviation.}
  \label{fig:taylor}
\end{figure}

Consider any smooth, positive function $H(z)$, with a local maximum at
$z=z_{\max}$. Close to this peak, Taylor's theorem implies some
(trivial) universality
\begin{equation}\label{eq:Taylor-universality}
  \frac{H(z)}{H(z_{\max})}=1-\frac{1}{2} y^2+{\cal O}(y^3)\,,
\end{equation}
where $ y=\sqrt{\frac{|H''(z_{\max})|}{H(z_{\max})}}(z-z_{\max})$.
Note that the peak's position is
$z_{\text{max}}$, its heigth is $H(z_{\max})$ and its (inverse) width is
$\sqrt{|H''(z_{\max})|/H(z_{\max})}$. In principle, there is no
reason for Eq.~\eqref{eq:Taylor-universality} to be accurate away from the
peak, but this formula suggests an alternative
representation for our $1-X$ curves, see Fig.~\ref{fig:taylor}. We note that,
in this new representation, the $1-X$ curves are invariant under changes of
the coherence length $\xi$ (upper panel). When
considering changes in the temperatures $T_1$ and $T_2$ and the probability
$F$, however, the curves mildly differ away from the peak (see Fig.~\ref{fig:taylor}
lower panel). This (approximate) independence of $(T_1,T_2,F,\xi)$ is a
fortunate fact because the complexity of the problem gets reduced to the study
of the scaling with $\xi$ of the three peak parameters while keeping
$(T_1,T_2,F)$ constant.

\begin{table}[t]
\begin{ruledtabular}
\begin{tabular}{cccccrc}
$F$ & $T_1$   & $T_2$  & $\xi_{\min}$ & $a$ & \multicolumn{1}{c}{$b$} &  $\chi^2/\mathrm{d.o.f}$ \\
\hline
0.001 & 0.625 & 0.7 &  4.75 &  0.60(12) & $0.9(9)$ &   22.12/19 \\
0.001 & 0.625 & 0.8 &  4.75 &  0.81(7) &  $0.0(5)$ &   11.52/19 \\
0.001 & 0.625 & 0.9 &  4.75 &  0.93(10) & $0.1(6)$ &   5.35/19 \\
0.001 & 0.625 & 1.0 &  4.75 &  1.13(13) & $-0.6(8)$ &  3.99/19 \\
\hline
0.001 & 0.7 &  0.8 &  5.00 &  0.88(9) &  $-0.6(7)$ & 38.18/27 \\
0.001 & 0.7 &  0.9 &  4.75 &  0.98(8) &  $-0.1(6)$ & 14.90/28 \\
0.001 & 0.7 &  0.0 &  4.75 &  1.08(7) &  $-0.2(6)$ & 22.32/28 \\
\hline
0.010 & 0.625 & 0.8 & 4.75 &  1.29(5) &  $-0.2(3)$ & 22.30/19 \\
0.010 & 0.625 & 0.9 & 4.75 &  1.47(6) &  $-0.5(4)$ & 7.32/19 \\
0.010 & 0.625 & 1.0 & 4.75 &  1.65(6) &  $-0.8(4)$ & 4.83/19 \\
\hline
0.010 & 0.7 & 0.8 &  5.25 &  1.23(7) &  $0.2(5)$ &   34.14/26 \\
0.010 & 0.7 & 0.9 &  4.75 &  1.48(9) &  $-0.7(6)$ &  17.19/28 \\
0.010 & 0.7 & 1.0 &  4.75 &  1.63(9) &  $-0.8(6)$ &  10.81/28 \\
\end{tabular}
\end{ruledtabular}
\caption{Parameters obtained in fits of our data for $N_{r,\max}^{1/3}$ to
Eq.~\eqref{eq:Nr_xi}. For each fit, we also report the figure of merit
$\chi^2/\mathrm{d.o.f}$ (we include in the fit data with $\xi \geq \xi_{\min}$
; $\xi_{\min}$ is set by requiring the fit’s $P$ value to be larger than
0.05).}
\label{tab:parametros_Nmax}
\end{table}
\begin{table}[t]
\begin{ruledtabular}
\begin{tabular}{c c c c c c c}
$F$ & $T_1$ &  $T_2$ &  $\xi{\min}$ &  $A$ & $\beta$ & $\chi^2/\mathrm{d.o.f}$ \\
\hline
0.001 &  0.625 &  0.7 &  4.75 & 0.8(3) & 0.9(2) &    18.72/19 \\
0.001 &  0.625 &  0.8 &  4.75 & 1.6(4) & 1.27(14) &  8.07/19 \\
0.001 &  0.625 &  0.9 &  4.75 & 1.4(3) & 1.32(12) &  10.05/19 \\
0.001 &  0.625 &  1.0 &  4.75 & 1.3(2) & 1.37(9) &   5.60/19 \\
\hline 
0.001 &  0.7 &   0.8 &  4.75 &  1.1(3) &   1.10(12) & 35.26/28 \\
0.001 &  0.7 &   0.9 &  4.75 &  1.26(16) & 1.25(7) &  25.90/28 \\
0.001 &  0.7 &   1.0 &  4.75 & 1.19(17) &  1.29(7) &  23.01/28 \\
\hline 
0.01 &  0.625 &  0.8 &  4.75 &  0.63(9) &   1.11(7) &   20.44/19 \\
0.01 &  0.625 &  0.9 &  4.75 &  0.59(10) &  1.14(8) &   6.08/19 \\
0.01 &  0.625 &  1.0 &  4.75 &  0.58(15) &  1.21(12) &  9.05/19 \\
\hline
0.01 &  0.7 &  0.8 & 4.75 & 0.59(11) & 1.05(12) &  21.26/28 \\
0.01 &  0.7 &  0.9 & 4.75 & 0.63(8) &  1.15(7) &   18.46/28 \\
0.01 &  0.7 &  1.0 & 4.75 & 0.59(12) & 1.18(9) &   17.93/28 \\
\end{tabular}
\end{ruledtabular}
\caption{Parameters obtained in fits of our data for $\kappa(\xi)$ to
Eq.~\eqref{eq:width_xi}. For each fit, we also report the figure of merit
$\chi^2/\mathrm{d.o.f}$ (we include in the fit data with $\xi \geq \xi_{\min}$
; $\xi_{\min}$ is set by requiring the fit’s $P$ value to be larger than
0.05).}
\label{tab:parametros_width}
\end{table}

\section*{Supplementary Note 5: Position and width of the peaks}
In this section we analyze the scaling of the peaks' position and 
width with the coherence length $\xi$.

We first focus on the peak's position, which signals the most
convenient length scale for studying TC (for a given coherence length $\xi$,
probability $F$ and temperatures $T_1$ and $T_2$). Dimensional analysis
suggests a linear fit as the natural ansatz to study the scaling of the
peak's position $N_{r,\max}^{1/3}$ with the coherence length $\xi(\tw)$
(indeed, both quantities are lengths): 
\begin{equation}
N_{r,\max}^{1/3} = a \> \xi(\tw) + b \, \, . \label{eq:Nr_xi}
\end{equation}
Fits of the data to Eq.~\ref{eq:Nr_xi} are shown in 
Table~\ref{tab:parametros_Nmax}. In all cases, values of parameter $b$ are
compatible with $0$ (at the two-$\sigma$ level). In addition, 
amplitude $a$ exhibits monotonic  behavior with $T_2-T_1$ and with the
probability $F$. Hence, our naive expectation $N_{r,\max}^{1/3} \propto
\xi(\tw)$ is confirmed.

The peak's width determines how delicate
the selection of the right length scale is to study TC (i.e., if the peak's
width becomes larger than its position, this choice is no longer critical).

We study the inverse peak's width (i.e., the curvature $\kappa(\xi)$) and
propose a power law decaying with $\xi(\tw)$ characterized by the ansatz
\begin{equation}
\kappa(\xi) = A(F) \, \xi^{-\beta} . \label{eq:width_xi}
\end{equation}
Results are shown in Table~\ref{tab:parametros_width}.

The value of $A(F)$ turns out to be compatible for all pairs of
temperatures $(T_1,T_2)$ at fixed probability $F$. Furthermore, at the current
precision of the data, exponent $\beta$ does not exhibit any significant
dependence either on the temperature pair ($T_1,T_2$) or on the probability $F$.

Let us now recall the linear relation between the peak's position and the
coherence length, see Eq.~\eqref{eq:Nr_xi}. Consider the ratio between the
position of the maximum and its width, $N_{r,\max}\kappa(\xi) \sim
\xi^{1-\beta}$. Table~\ref{tab:parametros_width} mildly suggest that
$\beta$ is slightly greater than 1, which implies that the ratio goes to zero
(very slowly) in the limit of large $\xi$.

\bibliographystyle{apsrev4-1_titles}

\end{document}